\documentstyle[epsfig, amssymb]{mn}  
  
\newcommand {\ds}{\displaystyle}  
  
\begin{document}  
\def\sun{\hbox{$\odot$}}  
  
\title{Strong lensing constraints on the velocity dispersion
and density profile of elliptical galaxies}  
  
\author[Davis, Huterer \& Krauss]{  
 Adam N. Davis$^1$, Dragan Huterer$^1$, and Lawrence M. Krauss$^{1,2}$  \\  
$ ^1$Department of Physics,  
 Case Western Reserve University,  
 Cleveland, OH~~44106 \\
$^2$ Department of Astronomy,  Case Western Reserve University,  
 Cleveland, OH~~44106}

\pubyear{2002}  
  
\maketitle  
  
\begin{abstract}  
We use the statistics of strong gravitational lensing from the
CLASS survey to impose constraints on the velocity dispersion and
density profile of elliptical galaxies. This approach differs from
much recent work, where the luminosity function, velocity dispersion
and density profile were typically {\it assumed} in order to constrain
cosmological parameters.  It is indeed remarkable that observational
cosmology has reached the point where we can consider using cosmology
to constrain astrophysics, rather than vice versa.  We use two
different observables to obtain our constraints (total optical depth
and angular distributions of lensing events). In spite of
the relatively poor statistics and the uncertain identification of
lenses in the survey, we obtain interesting constraints on the
velocity dispersion and density profiles of elliptical galaxies.  For
example, assuming the SIS density profile and marginalizing over other
relevant parameters, we find $168 \,{\rm km/s}\leq \sigma_* \leq
200\,{\rm km/s}$ (68\% CL), and $158 \,{\rm km/s}\leq \sigma_* \leq
220\,{\rm km/s}$ (95\% CL). Furthermore, if we instead assume a
generalized NFW density profile and marginalize over other parameters,
the slope of the profile is constrained to be $1.50\leq\beta\leq 2.00$
(95\% CL). We also constrain the concentration parameter as a function of
the density profile slope in these models.  These results are
essentially independent of the exact knowledge of cosmology. We
briefly discuss the possible impact on these constraints of allowing
the galaxy luminosity function to evolve with redshift, and also
possible useful future directions for exploration.
  
\vspace{1cm}  
\end{abstract}  
  
\section{Introduction}  
The statistics of strong gravitational lensing has repeatedly been
advertised and used as a probe of cosmology (e.g., Turner, Ostriker \&
Gott 1984, Hinshaw and Krauss 1987, Fukugita et al.\ 1992, Krauss and
White 1992, Kochanek 1995, 1996, Cooray, Quashnock \& Miller 1999,
Chiba \& Yoshii 1999,  Cheng \& Krauss 1999). 
The sensitivity of lensing counts to $\Omega_M$
and $\Omega_{\Lambda}$, the energy densities in matter and the vacuum
component relative to the critical, comes mostly from a volume effect:
higher $\Omega_{\Lambda}$ implies bigger comoving volume for a fixed
redshift, leading to the higher optical depth for lensing. Using
knowledge about the luminosity function of galaxies and their density
profiles, many authors have used lensing statistics to constrain
cosmological parameters. For example, Fukugita and Turner (1991) first
constrained the vacuum energy density to be less than about 90\% of
the critical energy density ($\Omega_{\Lambda}\lesssim 0.9$) at 95\%
confidence level (hereafter CL).  Subsequently this was followed by
Kochanek (1995, 1996), who claimed an upper limit on the vacuum energy
density ($\Omega_{\Lambda}<0.66$ at 95\% CL). Krauss and White (1992)
and later Chiba and Yoshii (1999) and Cheng and Krauss (1999) used a
different choice of galaxy parameters and demonstrated that a flat
vacuum-energy-dominated universe could be favored. Similar analyses
have been performed by Im et al.\ (1997), Cooray, Quashnock \& Miller
(1999), Waga \& Miceli (1999), and all typically favor the
$\Lambda$CDM cosmology.  Cheng and Krauss (1999, 2001) also explored
how uncertainties in the choice of galaxy parameters could result in
vastly different constraints on cosmology, although they argued for a
choice that ultimately favored a flat, vacuum-energy-dominated
cosmology.  It has also recently been argued that strong-lensing
statistics from ongoing surveys like the Sloan Digital Sky Survey
(SDSS) might impose interesting constraints on the equation-of-state
ratio of dark energy $w$ (Cooray \& Huterer 1999); constraints on $w$
from lensing have already been claimed by Sarbu, Rusin \& Ma (2001)
who used the statistics of the \ JVAS/CLASS survey to obtain
$w\lesssim -0.4$.  Similar results have been obtained very recently by
Chae et al. (2002).

Given the notoriously poor statistics of strong lensing surveys thus
far---the total number of gravitational lenses is of order fifty, and
the largest homogeneous survey (which we use in this work),
JVAS/CLASS, currently has a total of only 17 events--- combined with
the existing galactic luminosity function uncertainties, it is not
clear how seriously one should take any constraints on cosmology
derived from strong lensing statistics.  To constrain cosmological
parameters using lensing statistics one has to deal with the strong
dependence of the results on the lens profile, the density dispersion
of galactic dark matter, the number density of galaxies as a function
of redshift, and observational effects due to magnification bias and
the selection function of the survey.

In this work, we exploit this sensitivity to reverse the traditional
methodology.  Since lensing statistics are, on the whole, much more
sensitive to astrophysical than cosmological parameters, we wish to
utilize existing surveys to probe the properties of lensing galaxies
rather than cosmology.  We are aided in this effort at this time
because independent probes of cosmological parameters have recently
converged rather tightly on a single cosmological model: a flat dark
energy dominated universe with $\Omega_{\rm DE} \approx 0.7$,
and $\Omega_M \approx 0.3$. As these parameters currently seem  to be
more tightly constrained that the galaxy parameters described above,
now seems an opportune time to use cosmology to constrain
astrophysics, rather than vice versa!

Some efforts along these lines have already been explored, as new and
better lensing data, especially the JVAS/CLASS survey, have appeared.
In particular, several investigations have been undertaken to constrain
the nature of galaxy clustering in the CDM paradigm.  Keeton (2001)
used the statistics of JVAS/CLASS lenses to indicate that CDM galaxies
are too concentrated to agree with the lensing statistics, while
Keeton \& Madau (2001) used the absence of wide-separation lenses in
the CLASS survey to impose an upper bound on the concentration of dark
matter halos.  Takahashi
\& Chiba (2001) consider lensing by both singular isothermal sphere (SIS) 
and Navarro-Frenk-White (NFW) profile galaxies, and find that the lack
of observed large-angle separation lenses indicates that the density
profile is not too steep ($\beta\lesssim 1.5$, with $\rho(r)\propto
r^{-\beta}$). Oguri, Taruya \& Suto (2001) obtained a similar result
by using the statistics of tangential and radial arcs.  Conversely,
Rusin \& Ma (2001) use the absence of detectable odd images to set a
constraint on the surface density of lensing galaxies, and conclude
that lenses cannot have profiles much shallower than an SIS
($\beta\gtrsim 1.8$). Wyithe et al. (2001) and Li
\& Ostriker (2002) considered lensing by objects with both SIS and generalized 
NFW (GNFW) density profiles. They computed optical depths, image
separations and magnification biases. In particular, Li \& Ostriker,
extending the earlier work of Keeton (1998) and Porciani \& Madau (2000)
argued that, in order to explain the large number of observed
small-separation lenses and the lack of large-separation events
(compared to predicted distributions for lensing by clusters), the
favored galaxy cluster profile seems to be the combination of SIS
(when $M\lesssim 10^{13}M_{\sun}$) and NFW (when $M\gtrsim
10^{13}M_{\sun}$).

Here we carry out a related analysis, with the aim of constraining the
nature of individual galaxies rather than clusters.  For this purpose
we shall assume the ``concordance'' values for the cosmological
parameters (e.g. Krauss 2000): $\Omega_M=1-\Omega_{\rm DE}=0.3$,
$w=-1$ and $h=0.7$, where $\Omega_M$ and $\Omega_{\rm DE}$ are energy
densities in matter and dark energy relative to critical, $w$ is the
equation of state ratio of dark energy, and $H_0=100\,h\, {\rm
km/sec/Mpc}$. We will show that our results are extremely weakly
dependent on the assumed cosmology (in particular, knowledge of
$\Omega_M$).
  
\section{The Data}\label{sec:data}  
           
Although more than 60 multiply imaged quasars and radio sources are
known, they come from different observations with different
sensitivities and selection functions, which makes an accurate
computation of the expected number of lenses very
difficult. Therefore, it is imperative to have data from a single
well-understood survey with information on the source population. In
this work we use the most complete homogeneous sample of lenses
provided by the Cosmic Lens All-sky Survey (CLASS; Myers et al. 2002,
Browne et al. 2002), which extended the earlier Jodrell-Very Large
Array Astrometric Survey (JVAS; Patnaik et al. 1992a, King et
al. 1999). CLASS is using the Very Large Array to image radio sources
with the flux density of between $30$ and $200 {\rm mJy}$; candidate
lensing events are followed up by Multi-Element Radio-Linked
Interferometer Network (MERLIN) and the NRAO Very Large Baseline
Array (VLBI). So far a total of about 16,000 sources were imaged by
JVAS/CLASS, with 22 confirmed lensing events. Of these, a subset of
8958 sources with 13 lenses forms a well-defined subsample suitable for
statistical analysis (Browne et al. 2002), and we use this subsample 
in our work. Table 1, essentially identical to Table 3 in Browne et al. (2002),
shows the lenses from the statistically controlled subsample. We have
added information about the identity of the lens, in particular whether
it is a spiral galaxy, an elliptical, or formed by more than one galaxy
(Chae 2002).
  
\begin{table*}
\begin{tabular}{ l l c c c l l}
\hline \hline
Survey & Lens &  $z_l$ & $z_s$ & $\theta$ & ID & Reference \\
\hline 
JVAS    & B0218+357 & 0.68 & 0.96 & 0.33 & s & Patnaik et al.\ 1993 \\
CLASS   & B0445+123 & 0.56 & ---  & 1.33 & ? & Argo et al.\ 2002 \\
CLASS   & B0631+519 & ---  & ---  & 1.16 & ? & Browne et al.\ 2002 \\
CLASS   & B0712+472 & 0.41 & 1.34 & 1.27 & e & Jackson et al.\ 1998	 \\
CLASS   & B0850+054 & 0.59 & ---  & 0.68 & ? & Biggs et al.\ 2002 \\
CLASS   & B1152+199 & 0.44 & 1.01 & 1.56 & ? & Myers et al.\ 1999	 \\
CLASS   & B1359+154 & ---  & 3.21 & 1.65 & ?, m & Myers et al.\ 1999 \\
JVAS    & B1422+231 & 0.34 & 3.62 & 1.28 & e & Patnaik et al.\ 1992b  \\
CLASS   & B1608+656 & 0.64 & 1.39 & 2.08 & e, m & Myers et al.\ 1995 \\
CLASS   & B1933+503 & 0.76 & 2.62 & 1.17 & e & Sykes et al.\ 1998	 \\
CLASS   & B2045+265 & 0.87 & 1.28 & 1.86 & ? & Fassnacht et al.\ 1999\\
JVAS    & B2114+022 & 0.32/0.59 & ---  & 2.57 & e, m &  Augusto et al.\ 2001\\
CLASS   & B2319+051 & 0.62/0.59 & ---  & 1.36 & e &  Rusin et al.\ 2001   \\
\hline\hline
\end{tabular}
\caption{Thirteen lensing events from the ``CLASS statistical 
sample'' of 8958 objects (adopted from Browne et al. 2002; see also
Chae 2002). ``ID'' stands for identification of the lens - whether it
is a spiral galaxy (s), an elliptical (e) or unknown (?); three lenses
consist of multiple galaxies (m).}
\label{table}
\end{table*}

It is well known that elliptical galaxies dominate the optical depth
for strong lensing by individual galaxies (e.g. Kochanek 1993b), and
as a result we concentrate on constraining their parameters here.
This effort is somewhat complicated by the fact that only six of the
CLASS lenses are clearly identified as ellipticals and one as a
spiral, while in other cases the identity of the lens is uncertain;
see Table 1. Furthermore, three events are due to more than one lens
galaxy. It is crucial to choose a subset of CLASS lenses that includes
elliptical galaxies only. It is clear that the number of ellipticals
is between 6 and 12, and that confirmed ellipticals outnumber spirals
in ratio 6:1.  The most likely value of the number of
ellipticals is therefore somewhere near 11. 
To compute the measured number of lenses, we chose
marginalize over the range between 6 and 12, with the gaussian
weighting centered at 11 and variance of $5$. However, as we later discuss,
the results are extremely insensitive to the exact choice of
weighting; the reason is that the statistics are much more sensitive
to the parameters we wish to constrain, the velocity dispersion and
density profile of elliptical galaxies. For the angular separation
test, we use only the four {\it single}\footnote{Multiple lens
deflectors obviously make different predictions from single galaxies,
and produce larger angular separations of images. We are interested in
splittings due to single elliptical galaxies only.} elliptical lenses
(B0712+472, B1422+231, B1933+503, B2319+051).  To test the robustness
of this test, we alternatively assume that all unidentified galaxies
are ellipticals as well, and use a total of 9 {\it single} non-spiral
lenses (the four above, plus B0445+123, B0631+519, B0850+054,
B1152+199 and B2045+265). As discussed later on, our results are
insensitive to the exact choice of this subset.

We wish to utilize three different observables to obtain our
constraints: the overall optical depth $\tau$ to a source at redshift
$z_s$, the differential optical depth as a function of angular
separation, and the differential optical depth as a function of lens
redshift. Unfortunately, the last of these tests is uncertain due to
possible incompleteness of the survey: higher-redshift lenses are more
difficult to measure due to their lower fluxes; while the source
redshifts are more easily measurable for objects very far away (mainly
quasars) and very close (mainly galaxies), and not ones at
intermediate distances. Because of these uncertainties, and because
the redshift test does not add much to our constraints, we decide not
to use the redshift-distribution test\footnote{Nevertheless, we have
checked that the results of the redshift-distribution test agree with
those of the other two tests. Furthermore, for the SIS case the
quantity $(1/\tau) (d\tau/dz_l)$ is independent of galaxy parameters,
and we used it to check that constraint on $\Omega_M$ and $w$ is
consistent with the adopted cosmological model $\Omega_M=0.3$,
$w=-1$.}

We are therefore left with two tests, the total optical depth
($\tau$-test) and angular separation ($d\tau/d\theta$-test). The
former test gives stronger constraints in both SIS and GNFW cases.
The latter test, in the SIS case, is independent of $z_s$ as long as
$z_s\gtrsim 0.2$; henceforth, the knowledge of $z_s$ is not necessary
and all single ellipticals (chosen as explained above) can be used for
this test. In the GNFW case the knowledge of $z_s$ is required for
this test, and when it is not available we use the mean redshift of
the measured sources, $z_s= 2$. (We have checked that the
results change negligibly if, instead of $z_s= 2$, we use the
histogram of the source distribution from Marlow et al. (2000), which
is centered at $z_s=1.27$ and has long tails.)  Finally, we use the
maximum lens separation $\theta_{\rm max}$ as an estimator of the
angular separation $\theta$. Although this estimator has been widely
used in the literature due to the fact that $\theta_{\rm max}$ are
readily available, we warn that the angle corresponding to the average
image radius fitted to a lens model, for example, would be a better
estimator. Nevertheless, we do not expect that using $\theta_{\rm
max}$ will significantly bias the results, given the limited current statistics. 
Moreover, since higher
$\sigma_*$ roughly corresponds to larger angular separations, our
results may only be biased to {\it higher} $\sigma_*$, strengthening
our conclusion that this parameter is smaller than previously quoted
in the literature.

In order to compute the {\it expected} optical depth for any given
model, it is crucial to know the redshifts of source quasars and
galaxies.  The redshift distribution of JVAS/CLASS source objects has
been discussed in Marlow et al.\ (2000), who spectroscopically
followed up 42 sources at William Herschel Telescope. Most of these
sources are quasars; with a significant admixture of galaxies at
$z\lesssim 1$. The mean redshift of this subsample is $\langle
z_s\rangle=1.27$ with an rms spread of 0.95. In this work we use the
full histogram distribution of the observed subsample of sources
(Fig.\ 2 in Marlow et al.\ 2000), and assume that the redshift
distribution of the subsample gives a good representation of the
overall redshift distribution. One has to be cautious, however, since
the lensed sources come from a fainter population than the ones in
Marlow et al. (2000), and may be at different redshifts. The validity
of this assumption has been examined by Chae (2002), who reviews
existing observations and finds that the redshift distribution is
expected not to change much at lower flux densities, corresponding to
lensed sources.

Finally, we will need to know a few other details regarding the
CLASS sample. The survey is complete at image separations
$0.3''<\theta<15''$ (Helbig 2000, Myers et al. 2002). All confirmed
JVAS/CLASS lenses have image separations $\theta<3''$.  The
distribution of sources as a function of the total flux density $S$ is
well described by the power law

\begin{equation}
\frac{dn}{dS}\propto S^{-\eta}
\label{eq:dndS}
\end{equation}  

\noindent where $dn$ is the number of sources observed in flux density 
interval $dS$. For JVAS/CLASS, $\eta\simeq 2.1$ (Rusin \& Tegmark
2001).

\section{Density profile}
  
There is good evidence that the density profiles of dark halos on
cluster scales depend on the halo mass (Keeton 1998, Wyithe
et al.\ 2001, Li and Ostriker 2002). For the less massive halos ($M\lesssim
10^{13}M_{\sun}$), SIS profiles are found to be adequate, while for
large-mass halos ($M\gtrsim 10^{13}M_{\sun}$) NFW profiles provides a
good fit.  This result is also expected from semianalytic models,
which show that objects smaller than $M_c\approx 10^{13}M_{\sun}$ are
subject to baryonic cooling, whereby baryons collapse to the center
thereby enormously increasing the central density and lensing
cross-section, and converting the shallow NFW profiles into the steep
SIS (Rix et al. 1997, Kochanek \& White 2001, Keeton 2001).
  
Galaxy clusters tend to lead to large lens separations and/or extended
arcs and arclets. Since
  
\begin{equation}  
\theta=1.271'{D_{ls}\over D_s}\left (  
{M\over 10^{15}h^{-1}M_{\sun}}\right )^{2/3}  
\left ({\rho_{\rm crit}\over\rho_{\rm crit, 0}}\right )^{1/3}  
\end{equation}  
  
\noindent  (Li \& Ostriker 2002) where all quantities   
except $\rho_{\rm crit, 0}$ are evaluated at $z=z_l$, we see that
$M\gtrsim 10^{13}M_{\sun}$ corresponds to $\theta\gtrsim 3''$.  If we
are interested in primarily lensing by individual galaxies rather than
clusters, we should concentrate on image separations substantially
smaller than this value.  In CLASS, all lensing events have
separations smaller than $3''$. We therefore conclude that the
CLASS lenses are due to individual galaxies and not clusters.

Furthermore, we assume smooth, spherically symmetric density
profiles. This assumption is widely used, and supported by the
findings that the subclumps do not greatly affect the total optical depth
for lensing (Flores, Maller \& Primack 1996) and that asphericity of
density profiles affects mostly the ratio of quads to doubles and not
the optical depth (Rusin \& Tegmark 2001).

Our goal in this paper is twofold. First, we would like to constrain
the galaxy velocity dispersion {\it assuming} the SIS profile. The SIS
profile has repeatedly been used in the past to constrain cosmological
parameters, assuming the Schechter function parameters and the galaxy
velocity dispersion to be known.  We would like to reverse this
process and see whether the previously-used $\sigma_*$ is still
favored now that we have good knowledge of cosmological parameters.

Second, we would like to constrain the density profile of elliptical
galaxies. As argued above, only the inner parts of lens galaxies (a
few tens of kiloparsecs from the center) are responsible for
CLASS events.  Moreover, as discussed in Sec.~\ref{sec:GNFW},
there is good evidence that cores of galaxies are small and can safely
be ignored. Therefore, it seems justified to adopt $\rho(r)\propto
r^{-\beta}$ and try to constrain $\beta$. We do this via the
generalized Navarro-Frenk-White profile, as described in
Sec.~\ref{sec:GNFW}.

\section{Modeling the lens: Singular Isothermal Sphere (SIS) profile}  
\label{sec:model_SIS}  

\subsection{Number density of lenses}  
  
Since we are interested in elliptical galaxies, we adopt the Schechter
luminosity function (Schechter 1976) which has repeatedly been shown
to be a good fit to the measurements\footnote{In order to compute
optical depths for generalized dark matter distributions on cluster
scales, many authors have assumed the Press-Schechter mass function
(Press and Schechter 1976).  Since we are interested in constraining
observational properties of elliptical galaxies, and since the lens
identification from the CLASS survey indicates that most lenses
are due to individual galaxies, for our purposes the Schechter
luminosity function is more relevant.}
  
\begin{equation}  
{d\phi\over dL}(L)\,dL=\phi_*\left (L\over L_*\right )^\alpha  
\exp(-L/L_*) {dL\over L_*}.  
\label{eq:sch_fun}
\end{equation}  
  
\noindent  There has been much discussion as to what values of
$\phi_*$ and $\alpha$ best describe the actual luminosity
function. Typically, it is argued that $\phi_{*, {\rm TOT}}=1.4\times
10^{-2}\,h^3\,{\rm Mpc}^{-3}$ for all galaxies, of which $\approx
30\%$ are ellipticals (Postman \& Geller, 1984), so that
$\phi_*^{ellip}=0.6\times 10^{-2}\,h^3{\rm Mpc}^{-3}$; further,
$\alpha\approx -1$ with fairly large uncertainties. Recently the SDSS
(Blanton et al. 2001) claimed a more accurate determination of the
local ($z\lesssim 0.2$) luminosity function; $\alpha=-1.20\pm 0.03$
and $\phi_{*, {\rm TOT}}=(1.46\pm 0.12)\times 10^{-2}\,h^3\,{\rm
Mpc}^{-3}$.
  
To relate the luminosities to velocity dispersions, we use the  
Faber-Jackson relation (Faber \& Jackson 1976)  
  
\begin{equation}  
\left (L\over L_*\right )=\left (\sigma\over \sigma_*\right )^{\gamma}  
\end{equation}  
  
\noindent where it is typically assumed that $\gamma\approx 4$ for 
the SIS profile. Our principal goal is to determine the parameters
$\phi_*$, $\alpha$, $\gamma$ and $\sigma_*$.

\subsection{Optical depth}  
   
  
The optical depth for a lens at redshift $z_l$ due to a  
particular source at $z_s$ is given by  
  
\begin{eqnarray}  
\tau &=& \int_0^{z_s} dz_l {dD_l\over dz_l}\, (1+z_l)^3\times\nonumber\\   
&&\int_0^\infty dL {d\phi\over dL}(L)\,\sigma_{\rm SIS}(L, z_l, z_s)
B(L, z_l, z_s)  
\label{eq:opt_depth}  
\end{eqnarray}  
  
\noindent where   
$\phi$ is the comoving number density of lenses, $L$ is their
luminosity, $\sigma_{\rm SIS}(L, z_l, z_s)$ is their cross-section for
lensing, and $B(L, z_l, z_s)$ is the magnification bias, describing
the fact that lensed galaxies will be magnified, and therefore seen
more easily, and therefore are enhanced in any flux limited survey. In
Eq.~(\ref{eq:opt_depth}) we have allowed for a general redshift and
luminosity dependence of the number density, cross-section and
magnification. For redshift-independent (as we first assume) $\phi_*$,
$\alpha$ and $\gamma$, $d\phi/dL$ depends only on $L$. Similarly,
assuming that CLASS lenses are described by an SIS profile and
the radio luminosity function is a power law, $B$ is simply a constant
(see below).
  
The density profile for the SIS is given by  
  
\begin{equation}  
\rho(r)={{\sigma}^2\over 2\pi Gr^2}  
\end{equation}  
  
\noindent where $\sigma$ is the velocity dispersion of the galaxy. This  
distribution produces an image separation of $2\theta_E$, where the
Einstein radius $\theta_E=4\pi(\sigma/c)^2D_{ls}/D_s$, and
$D_{ls}$ and $D_s$ are the angular diameter distances between the lens
and source, and observer and source respectively. The cross-section
for lensing is therefore
  
\begin{equation}  
\sigma_{\rm SIS}=\pi(\theta_E D_l)^2=16\pi^3  
\left (\sigma\over c\right )^4\,\left (D_l D_{ls}\over D_s\right )^2  
\label{eq:SIS_cross_sec}  
\end{equation}

\begin{figure*}
\includegraphics[height=3.7in, width=2.8in, angle=-90]
	{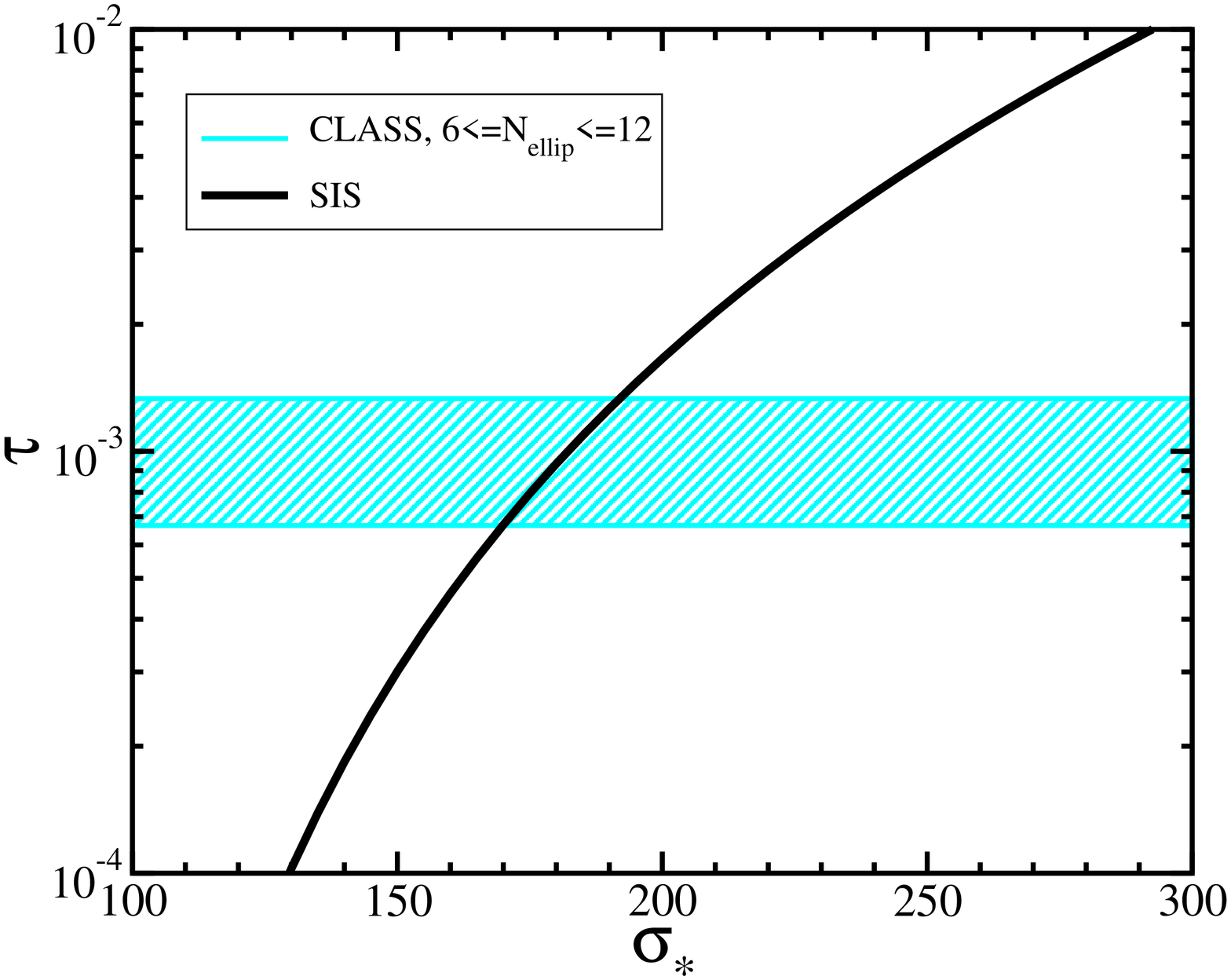}\nobreak\hspace{-0.4cm}
\includegraphics[height=3.7in, width=2.8in, angle=-90]
	{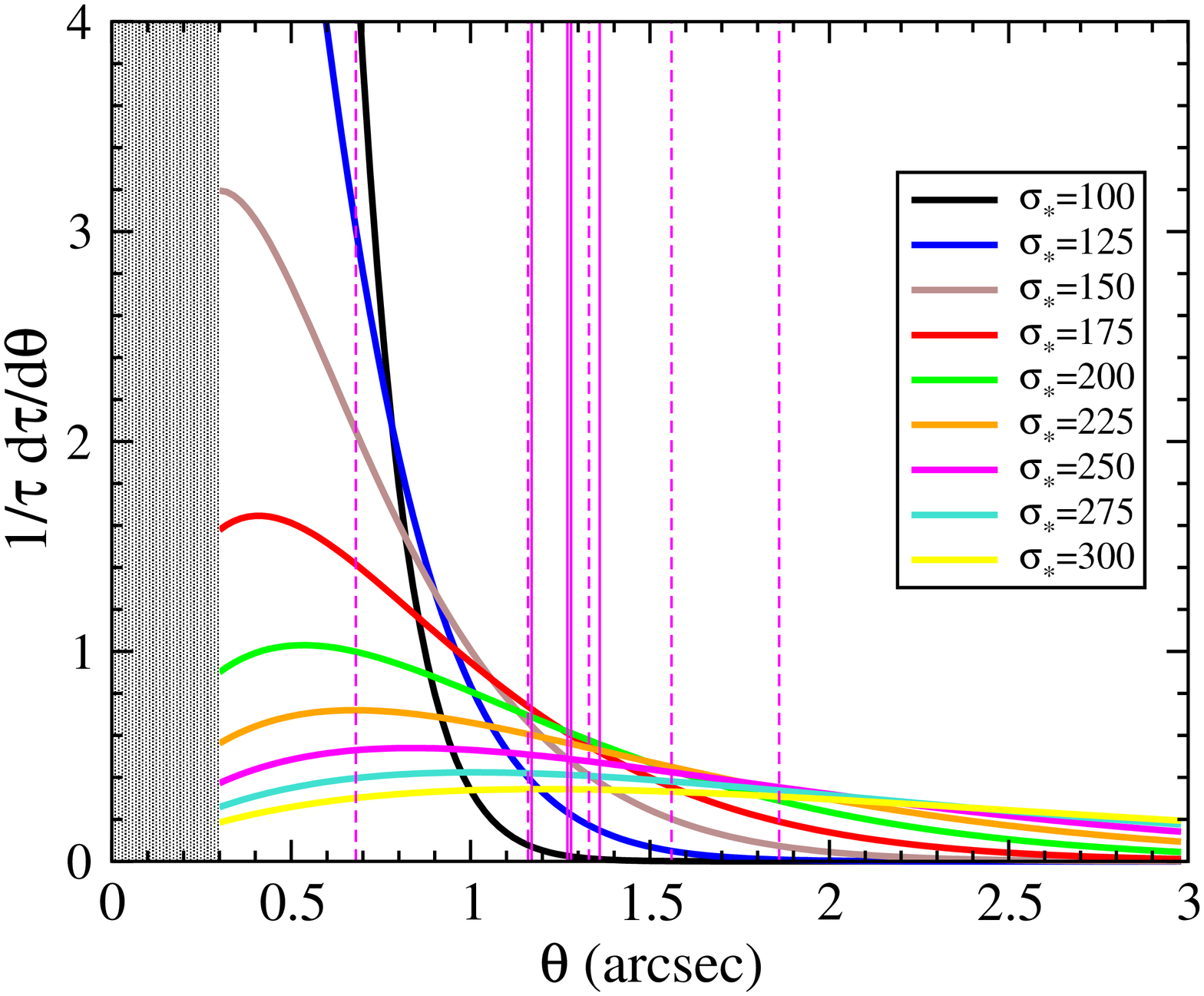}
\caption{Dependence of the observables on the velocity dispersion $\sigma_*$,
assuming that all other parameters take their fiducial values.  Left
panel: The dependence of $\tau$ (the shaded region is the measured
value from CLASS, assuming the number of ellipticals to be between 6
and 12). Right panel: The dependence of $(1/\tau) (d\tau/d\theta)$
(vertical lines denote measurements from CLASS, solid lines denote
confirmed ellipticals, while dashed lines denote galaxies whose type
has not been identified). The other Schechter-function and
cosmological parameters were fixed to their fiducial values from
Sec.~\ref{sec:SIS_depend}.  Note that the CLASS survey is complete for
$\theta>0.3''$; therefore, all predicted quantities, such as
$d\tau/d\theta$ in the right panel, were compared to measurements only
for $\theta>0.3''$. The quantity $d\tau/dz_l$ is very weakly dependent
on $\sigma_*$ (and other Schechter function parameters) and is not
shown.}
\label{fig:SIS_depend}
\end{figure*}

We only consider angular separations greater than some minimum value
$\theta_{\rm min}$, since the resolution limit of CLASS is
$\theta_{\rm min}=0.3''$, and multiple images with smaller separation
angles than $\theta_{\rm min}$ will not be resolved. The
correspondence between the luminosity and angular separation for an
SIS lens is
  
\begin{equation}  
L=\left ({\theta\, D_s\, c^2\over 8\pi\, D_{ls}\, {\sigma}_*^2}  
\right )^{\gamma/2}\,L_*  
\label{eq:L_vs_theta}  
\end{equation}  
  
\noindent where $c$ is the speed of light, so that $\theta_{\rm min}$   
corresponds to some $L_{\rm min}$ as the lower limit of integration in  
Eq.~(\ref{eq:opt_depth}).  

We also need to compute the magnification bias. It is given by

\begin{equation}
B  = \frac{\ds\int {dn\over dS}\, \ds{S\over \mu}\, P(\mu) \mu^{-1} d\mu}
		{\ds\frac{dn}{dS}}
\end{equation}

\noindent where $dn/dS$ is the source luminosity function and $P(\mu)$ the
distribution of total magnifications. For the power-law luminosity
function of CLASS (cf. Eq.~(\ref{eq:dndS})) and the distributions
of magnifications for SIS lenses ($P(\mu)=8\mu^{-3}$), the bias
simplifies to (Sarbu, Rusin \& Ma 2001)

\begin{equation}
B(L, z_l, z_s)=4.76.
\end{equation} 

Finally, we will be interested in the quantity $d\tau/d\theta$. This is
given by
  


\begin{eqnarray}  
\left.{d\tau\over d\theta}\right|_{\theta_1}  
&=& \left.{d\tau\over d\L}\right|_{L(\theta_1)}\times  
\left.{d\L\over d\theta}\right|_{\theta_1} \nonumber\\[0.1cm]  
&=&\int_0^{z_s} {dD_l\over dz}(z_l)\, dz_l\,(1+z_l)^3  
{d\phi\over dL}(L, z_l)\, \times \\[0.1cm] 
&& \sigma_{\rm SIS}(L, z_l, z_s)\, 
B(L, z_l, z_s)\times\left.{d\L\over d\theta}\right|_{\theta_1}\nonumber,  
\end{eqnarray}  
 
\noindent and the correspondence between $L$ and $\theta$ is given by  
Eq.~(\ref{eq:L_vs_theta}) Note also that we have allowed, in these
formulas, for a general dependence of the luminosity function, $L$, on
$z_l$, which we shall consider later in this paper.

\subsection{Dependence on parameters} \label{sec:SIS_depend} 
 
To illustrate the dependence of our observables ($\tau$  and 
$d\tau/d\theta$) upon the parameters, we assume for a
moment the following fiducial values: $\phi_*=0.6\times
10^{-2}\,h^3{\rm Mpc}^{-3}$, $\alpha=-1$, $\gamma=4$ and
$\sigma_*=180\,{\rm km/s}$.  For purposes of this illustration we have
also assumed all sources to be at a fixed redshift, chosen to be
$z_s=1.3$.

The dependence of lensing statistics on the galaxy parameters and
various degeneracies between these parameters have been investigated
extensively in the literature (see e.g. Kochanek 1993a, 1993b); here
we present a brief overview.  The variation of the total optical depth
$\tau$ around this fiducial model can easily be computed to be
  
\begin{eqnarray}  
d\ln{\tau} &=& 2.07\,d\ln{z_s}+1.00\,d\ln{\phi_*}+0.69\,d\ln{\alpha}+
	\nonumber\\  
&& 4.16\,d\ln{\sigma_*} + 0.69\,d\ln{\gamma}-0.61\,d\ln{\Omega_M}+\nonumber\\
 && 0.61\,d\ln{w}.  
\label{eq:depend}
\end{eqnarray}  
  
Perhaps not surprisingly, the strongest dependence is on the velocity
dispersion, which strongly affects the lensing cross-section, as well
as the luminosity function. $\phi_*$ enters linearly, and is
degenerate with other factors, for example the magnification bias
which is also a pure constant in the SIS case. Note, however, the much
weaker dependence upon the cosmological parameters $\Omega_M$ and $w$.
This reinforces the notion, independent of observational
uncertainties, that lensing constraints might most effectively be used
to constrain galaxy profile and luminosity function parameters, in
particular $\sigma_*$, rather than cosmological parameters.
  

Fig.~\ref{fig:SIS_depend} shows the dependence of $\tau$ and
$(1/\tau)(d\tau/d\theta)$ on $\sigma_*$. As expected, $\tau$ is a
strongly increasing function of $\sigma_*$, while
$(1/\tau)(d\tau/d\theta)$ favors higher angular splittings with
increasing $\sigma_*$.  As we shall describe, the fact that we only
compare theory with observation for $\theta>0.3''$ (the 
angular resolution of the survey) allows the likelihood function for
angular splitting to be consistent with that for optical depth, which
favors models with low $\sigma_*$.

We briefly comment on the dependence on other parameters. Assuming
$\phi_*={\rm const}$, only $\tau$ depends on this quantity (we show in
Sec.~\ref{sec:evolving} that this is essentially true even if $\phi_*$
is redshift-dependent). Since $\tau$ scales directly with $\phi_*$,
the presence of other parameters implies that constraints on $\phi_*$
will be very weak. Furthermore, it is clear that, in the SIS case,
$\tau$ only depends on the combination $\alpha+4/\gamma$ (this is
slightly spoiled by the fact that the luminosity integral starts at
$L_{\rm min}>0$). We found that even trying to constrain this combination 
gives weak constraints -- from either $\tau$ or $d\tau/d\theta$ test. The
only parameter that we are able to significantly constrain is $\sigma_*$.

\section{The likelihood function}  
  
As we have mentioned there are several ways to use statistics of
strong gravitational lensing.  The total number of lenses -- predicted
vs. observed -- is an obvious and most commonly used statistics which
provides information about the integrated optical depth for lensing.
The angular splitting and redshift distribution of lenses are the
other statistical observables, and in this work we use the former. We
choose not to use the redshift distribution due to selection effects
that are presumed to be significant in this test. Nevertheless, we
have checked that the redshift distribution, if included,  adds results
consistent with the other two constraints.

The probability of the total optical depth can be computed using the
Poisson distribution (e.g.\ Kochanek 1993b)
  
\begin{equation}  
{{\mathcal{L}}}_{\tau} = {N^x \exp(-N)\over x!}  
\end{equation}  
  
\noindent where  $x$ is the number of adopted lenses in the
CLASS survey, $N=8958\tau$ is the number of galaxies predicted
by the model, and $\tau(\phi_*, \sigma_*, \alpha, \gamma)$ is the
computed optical depth given the Schechter function and cosmological
parameters.  This formula gives the correct likelihood for any value
of $\tau$.  Recall that our determination of $\tau$ was based on the
Marlow et al. (2000) subsample redshift distribution.

The likelihood for the angular distribution of galaxies
is
 
\begin{equation}  
{\mathcal{L}}_{d\tau/d\theta} = \prod_{i=1}^{M} \,{1\over \tau_i}\,  
	\left.{d\tau\over d\theta} \right|_{\theta_i},    
\end{equation}  

\noindent where the product runs over the $M$ lenses which we want to use 
for this test (recall, we use alternatively $M=4$ or $M=9$, and get
virtually identical results for the two cases).

Finally, the joint likelihood for the redshift and angular distribution of
galaxies, which takes into account correlations between these two
observables, is given by
  
\begin{equation}  
{\mathcal{L}}_{d^2\tau/dzd{\theta}} = 
	\prod_{i=1}^{6} \, {1\over \tau_i}\,\left.{d^2\tau\over dzd{\theta}}  
		\right|_{z_l(i)\theta_i}.   
\end{equation}  
  
\noindent As mentioned in Sec.~\ref{sec:data}, we do not quote results from 
this test due to uncertainties regarding the redshift completeness.   We do
 illustrate the constraints it gives in the GNFW case to demonstrate that including
 this result would not change our conclusions.

The total likelihood we use is

\begin{equation}
{\mathcal{L}}_{\rm TOT}={\mathcal{L}}_{\tau}
	\times {\mathcal{L}}_{d\tau/d\theta}
\end{equation}

\noindent and it depends on cosmological parameters, as well as the Schechter
function parameters $\phi_*$, $\alpha$, $\gamma$ and $\sigma_*$.

Equation~(\ref{eq:depend}) suggests that, for the SIS profile, by far
the strongest dependence amongst the various lensing statistics is on
the velocity dispersion $\sigma_*$.  We determine the likelihood of
$\sigma_*$ by marginalizing over the other parameters:

\begin{equation}
{\mathcal{L}}(\sigma_*)=\int {\mathcal{L}}(\sigma_*, \phi_*, \alpha, \gamma)
	\, d\phi_*\,d\alpha\, d\gamma.
\end{equation}

\noindent where ${\mathcal{L}}$ refers to any combination of the 
likelihood functions discussed above.

\section{SIS profile -- Results}\label{SIS:results}

\begin{figure*}
\includegraphics[height=3.5in, width=2.5in, angle=-90]
	{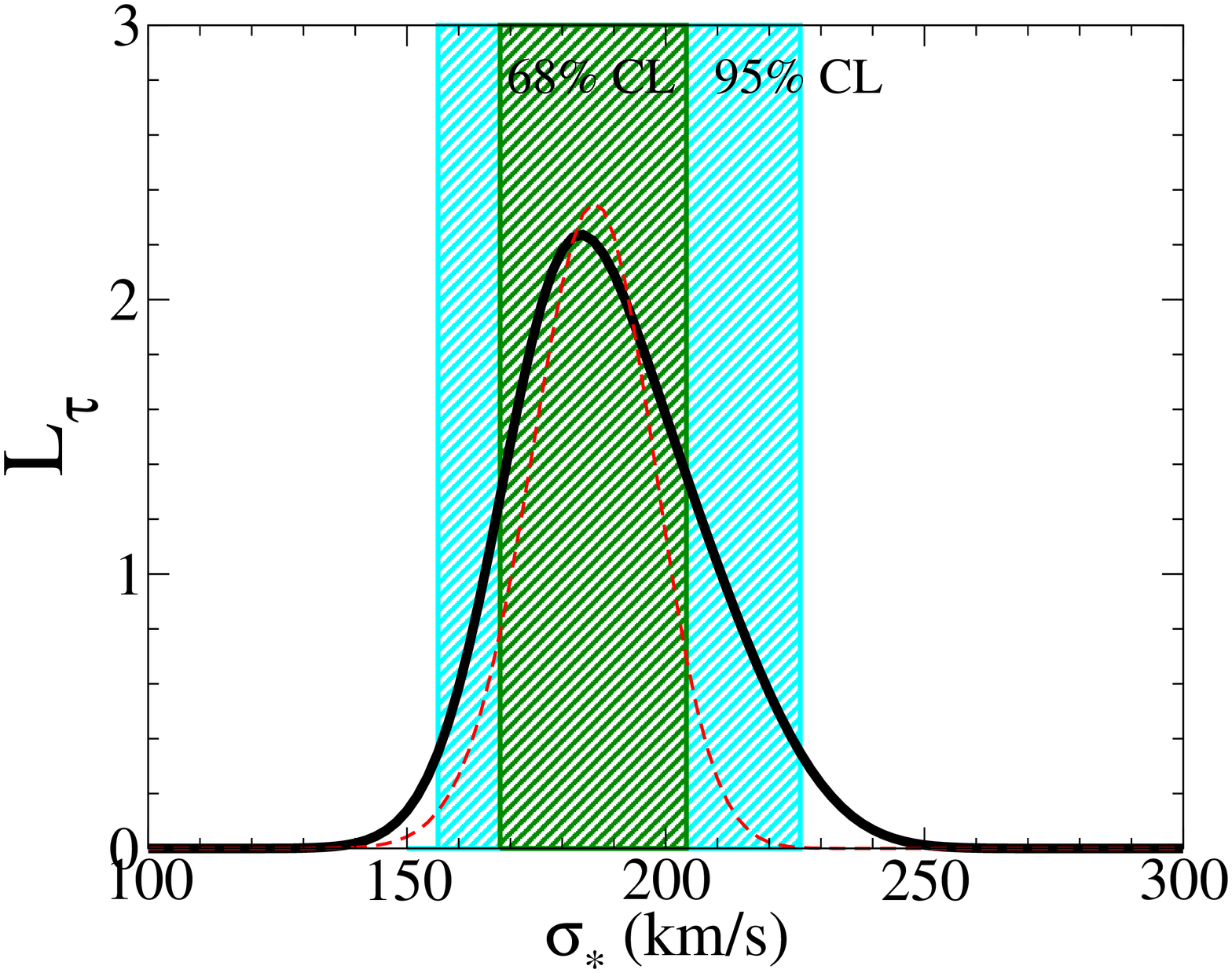}\nobreak\hspace{-0.4cm}
\includegraphics[height=3.5in, width=2.5in, angle=-90]
	{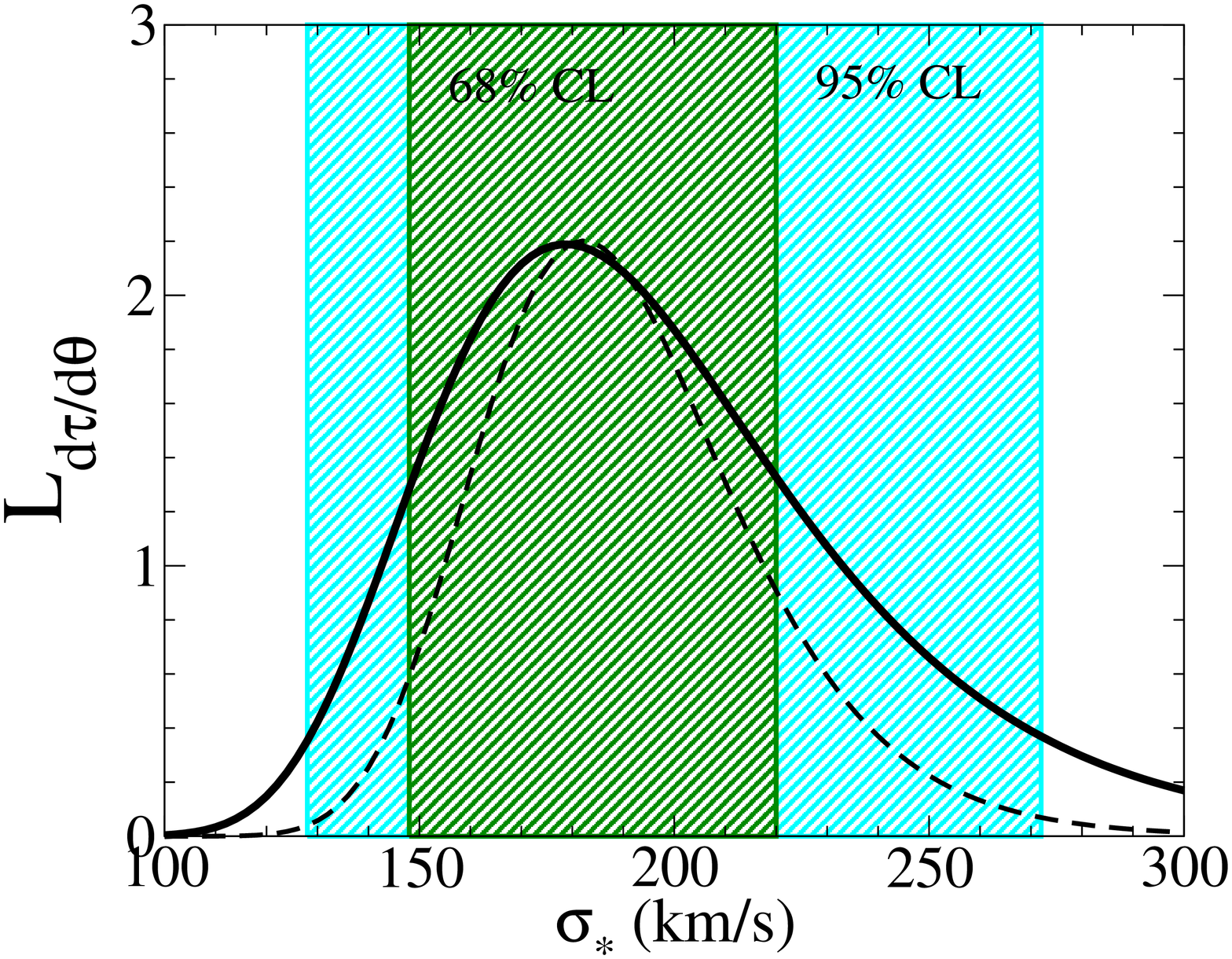}\\[-0.15cm]
\includegraphics[height=3.5in, width=2.5in, angle=-90]
	{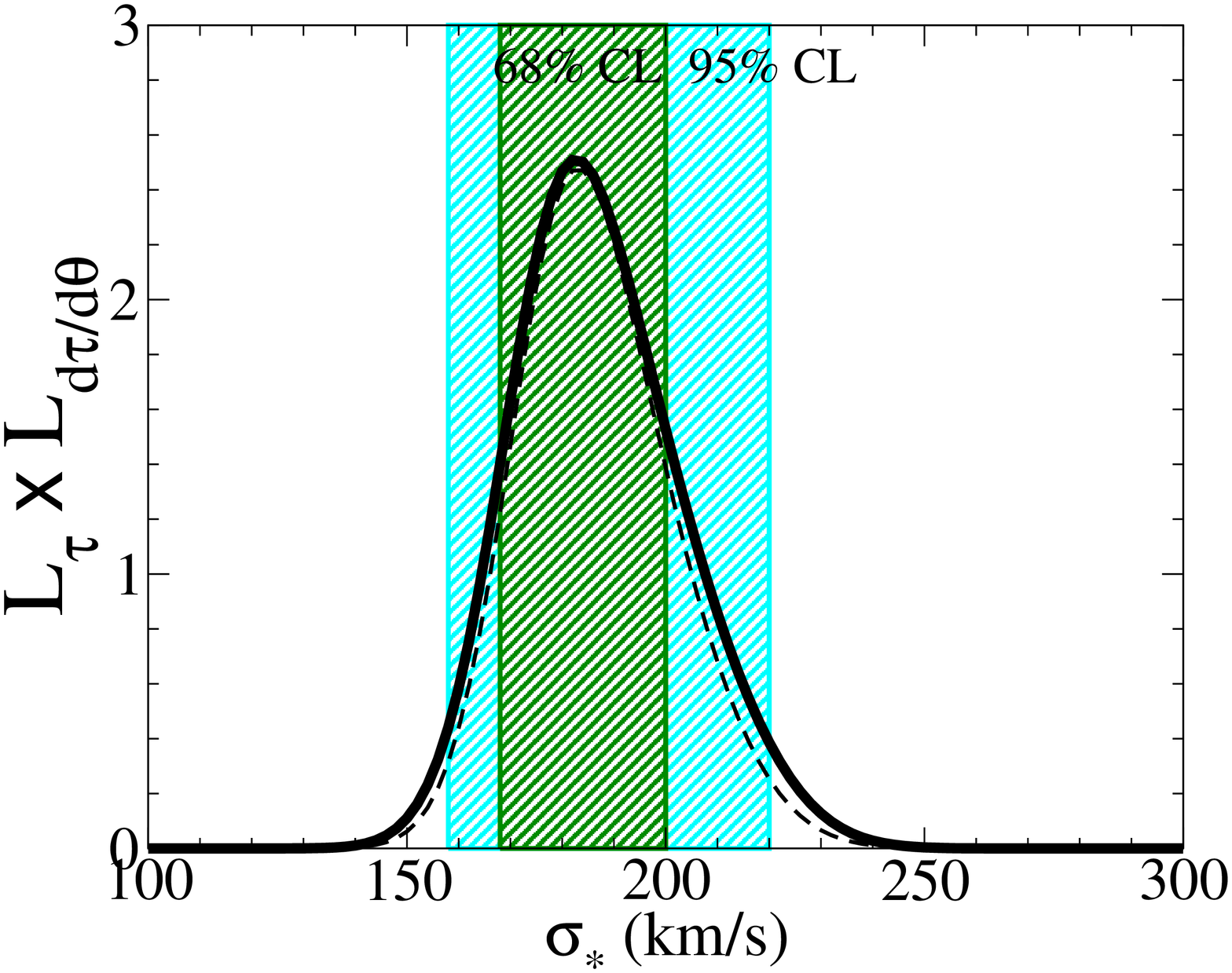}\nobreak\hspace{-0.4cm}
\caption{Constraints on the velocity dispersion $\sigma_*$ assuming the SIS 
lens profile and marginalizing over the other luminosity-function
parameters. Top left: Constraint from the $\tau$-test (for comparison,
the red-dashed line denotes the case when the galaxy parameters have
been fixed to their fiducial values of $\phi_*=0.6\times 10^{-2} h^3
{\rm Mpc}^{-3}$, $\gamma=4.0$, $\alpha=-1.0$). Top right: constraint
from the $d\tau/d\theta$-test, assuming 4 events that are due to
ellipticals (solid line) and additional 5 events that are due to
unidentified galaxies (dashed line). Bottom: the two tests combined;
the two curves refer to the two subsamples used in the $d\tau/d\theta$
test. Our baseline results, which we quote and to which the shaded
confidence regions correspond, refer to solid curves in the three
panels.}
\label{fig:SIS_constr}
\end{figure*}
  
As mentioned above, the strong dependence of the optical depth on the
velocity dispersion $\sigma_*$ implies that we might hope to get an
interesting constraint on $\sigma_*$ despite the relatively poor
lensing statistics and degeneracies between lensing parameters. We
marginalize over the other three relevant parameters, which we give
top-hat (uniform) priors of $\phi_*\in [0.5, 1.5]\times 0.6\times
10^{-2} h^3 {\rm Mpc}^{-3}$, $\gamma\in [3.0, 4.0]$, $\alpha\in [-1.3,
0.7]$. These ranges are conservative, allowing the full spread of
values reported in various recent measurements.  We also made sure to
use intervals that are {\it symmetric} around the traditionally
favored values, although it turns out that the exact choice of
intervals affects the results very weakly. For example, the SDSS, from
its commissioning data (Blanton et al. 2001), indicates that
$\alpha=-1.20\pm 0.03$ and $\phi_{*, {\rm TOT}}=(1.46\pm 0.12)\times
10^{-2}\,h^3\,{\rm Mpc}^{-3}$, while the Two Degree Field survey from
their preliminary sample of 45000 galaxies (Cross et al. 2001) gives
$\alpha=-1.09\pm 0.03$ and $\phi_{*, {\rm TOT}}=(2.02\pm 0.02)\times
10^{-2}\,h^3\,{\rm Mpc}^{-3}$; both of these quote the {\it total}
luminosity function. Kochanek et al.\ (2001), on the other hand,
isolated early type galaxies from the K-band luminosity function,
obtaining $\alpha=-0.92\pm 0.10$ and $\phi_{*}=(0.45\pm 0.06)\times
10^{-2}\,h^3\,{\rm Mpc}^{-3}$.  Finally, there are direct, independent
constraints on the Faber-Jackson slope $\gamma$ from the lens data; 
for example, Rusin et al.\ (2002) find $\gamma=3.44\pm 0.58$ for
early-type galaxies.

Fig.~\ref{fig:SIS_constr} shows the 68\% and 95\% CL constraints from 
the $\tau$ and $d\tau/d\theta$ tests (top panels), as well as the
constraints from the two tests combined (bottom panel). These
constraints correspond to solid curves in the three panels; for
comparison, the dashed line in the first panel we show the effect of
fixing $\phi_*$, $\alpha$ and $\gamma$ to their ``fiducial'' values,
while the dashed lines in the top right and bottom panel indicate the
effect of including the galaxies that are not identified as
ellipticals in the angular separation test. First of all, note that
the two independent tests are in remarkable agreement, and that both
constrain $\sigma_*$ quite strongly.  The $\tau$ test gives $156
\,{\rm km/s}\leq \sigma_* \leq 226\,{\rm km/s}$ (at the 95\% CL),
while the $d\tau/d\theta$ test gives $128 \,{\rm km/s}\leq \sigma_*
\leq 272\,{\rm km/s}$ (95\% CL). Moreover, the $d\tau/d\theta$ results 
are roughly independent of the subsample of ellipticals we use,
although the results are less tight in the baseline case when only the
four ``secure'' ellipticals are used; see Fig.~\ref{fig:SIS_constr}. 
The two tests combined give $158
\,{\rm km/s}\leq \sigma_* \leq 220\, {\rm km/s}$ (95\% CL).
Therefore, the overall favored value of $\sigma_*$ is actually smaller
than the fiducial value of $225
\,{\rm km/s}$ that has often been used to set constraints on
cosmological parameters (Kochanek 1995, 1996, Falco, Kochanek \&
Mu\~{n}oz 1998, Waga \& Miceli 1999, Cooray, Quashnock \& Miller
1999), and, not surprisingly, is in agreement with the value used in
studies that tended to favor non-zero $\Lambda$ (Cheng and Krauss,
1999, Chiba and Yoshii 1999). Note, however, that $\sigma_*\approx 225
\,{\rm km/s}$ has also been obtained using the direct observations of
early-type lens galaxies (e.g. Koopmans \& Treu 2002 get
$\sigma_*\approx (225\pm 15) \,{\rm km/s}$). Our results disfavor this result as representing
a fiducial value.

We found that the constraint on $\sigma_*$ is very weakly dependent on
the exact value of intervals allowed for other
parameters. Furthermore, we find that independent constraints on other
parameters of interest ($\phi_*$, $\alpha$ and $\gamma$) are very
weak, as expected from Eq.~(\ref{eq:depend}) and the fact that these
parameters are highly correlated (e.g. $\alpha$ and $\gamma$).
Finally, we have checked that the dependence of these results on
cosmology is extremely weak: for example, marginalizing over the
plausible values of the matter density $\Omega_M\in [0.15, 0.40]$
(while maintaining the flatness condition) produces likelihoods that
are only slightly broader.

\section{Modeling the lens: GNFW profile}\label{sec:GNFW}  
  
There is a good evidence that galaxies have a cuspy inner
profiles. The strongest argument comes from N-body simulations, which
argue for a profile $\rho(r)\propto r^{-\beta}$ with $\beta\simeq 1$
(Navarro, Frenk \& White 1996, 1997) or perhaps $\beta\simeq 1.5$
(Moore et al.\ 1999, Ghigna et al.\ 2001) -- in either case, a
relatively steep profile.  Another argument in favor of strongly
cusped central profiles is given by the absence of central images in
CLASS; assuming $\rho(r)\propto r^{-\beta}$ one obtains
$\beta>1.8$ at 95\% CL (Rusin \& Ma 2001). Finally, direct modelling
of the observed lenses favors steep inner cusps with profiles close to
isothermal; $\rho(r)\propto r^{-2}$ (Mu\~{n}oz, Kochanek \& Keeton
2001, Cohn et al.\ 2001, Treu \& Koopmans 2002, Winn, Rusin \&
Kochanek 2002). These and other lines of evidence suggest that the
central profiles of lens galaxies are steep and that cores, if they
exist, are tiny, with radius of a few tens or hundreds of parsecs at
most. Such small cores would not affect the lensing observables
appreciably (Hinshaw \& Krauss 1987).
  
To attempt to constrain the detailed profiles of elliptical galaxies  
we must move beyond the simple SIS model.  
In order to explore the dependence of lensing statistics on the details
of the density  
profile, we adopt the generalized NFW 
profile described below.

\subsection{The GNFW profile}  
  
The generalized NFW (GNFW) profile (Zhao 1996) is given by 
  
\begin{equation}  
\rho(r) = {\rho_s \over \left (\ds{r\over r_s}\right )^\beta   
\left [1+\left (\ds{r\over r_s}\right )\right ]^{3-\beta}}  
\end{equation}   
  
\noindent where $r_s$ is the characteristic scale where the density profile
shape can change.  Because the integral of this density profile
diverges at infinity, the mass of the halo is defined to be the mass
contained within the radius $r_{200}$ at which the density is 200
times greater than the critical density of the universe {\it at that redshift}:

\begin{equation}
M \equiv M_{200} = 200\left ({4\pi\over 3}r_{200}^3(z)\rho_c(z)\right )
\label{eq:M200_1}
\end{equation}

The expression for the mass can further be written as
  
\begin{equation}  
M = 4\pi\int_0^{r_{200}(z)}\rho r^2dr = 4\pi\rho_s(z) r_s^3(z)\,f(c(z))  
\label{eq:M200_2}
\end{equation}  
  
\noindent where 

\begin{equation}  
f(c) \equiv \int_0^{c}  
\frac{x^2dx}{x^\beta(1+x)^{3-\beta}}.  
\label{eq:f}
\end{equation}  

\noindent and the concentration parameter is defined as  
  
\begin{equation}  
c(z) \equiv {r_{200}(z)\over r_{s}(z)}.  
\label{eq:c}
\end{equation}

From Eqs.~(\ref{eq:M200_1})-(\ref{eq:c}) it follows that  
  
\begin{eqnarray}  
r_s(z)    &=& \frac{1}{c(z)}\,
	\left (\frac{3M_{200}}{800\pi \rho_c(z)}\right )^{1/3}	
	\label{eq:rs}\\[0.15cm]  
\rho_s(z) &=& \frac{200}{3}\rho_c(z)\frac{c(z)^3}{f(c(z))}.  
\label{eq:rhos}
\end{eqnarray}  
  
Thus, the generalized NFW profile is determined by the choice of
the inner density slope $\beta$ and the concentration $c(z)$. Starting
with these two parameters, one can compute $\rho_s(z)$ from
Eq.~(\ref{eq:rhos}) and then, given the mass of the halo, $r_s(z)$
from Eq.~(\ref{eq:rs}). Note that the GNFW profile for $\beta=2$ and
the SIS profile are different for three reasons: 1) the GNFW profile
parameters are explicitly redshift-dependent, 2) the two profiles have
different normalizations, and 3) the GNFW profile has a turnover at
$r=r_s$, while the SIS does not.
  
\subsection{The halo concentration}  

The halo concentration factor $c(z)$ is fortunately fairly well
constrained due to recent results obtained using N-body simulations
(e.g. Bullock et al.\ 2001a, Wechsler et al.\ 2002). For a pure NFW
profile, the concentration of the halos is well described by
  
\begin{equation}  
c(z)={c_0\over (1+z)} \left ({M\over M_*}\right )^{-0.13}  
\label{eq:c_Nbody}  
\end{equation}  
  
\noindent with $c_0=9$ and $M_*=1.5\times 10^{13}M_{\sun}$ 
(the above papers actually quote results for $c_{vir}\equiv r_{\rm
vir}/r_s$ with $r_{\rm vir}$ a virial radius, but the formula we quote
accounts for the difference in definition quite accurately). The
dependence on $M$ is small and does not change the results much, while
the dependence on redshift is important and fairly well-understood
(Wechsler et al.\ 2002). It is also important to account for the
variance in $c$ which occurs not only because of uncertainties in halo
modelling, but also because of the variance in halo properties. We
adopt an uncertainty in $\log_{10}c_0$ to be 0.14 (Bullock et al.\
2001a, Wechsler et al.\ 2002). Therefore, when computing the likelihood
function we weight excursions around the middle value of $c_0$ by a
gaussian factor with this standard deviation.
  
\begin{figure}  
\includegraphics[height=3.5in, width= 3in, angle=-90]{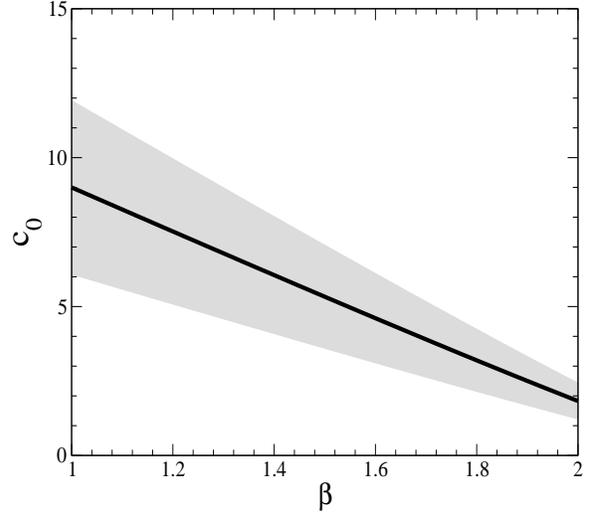}  
\caption{The mean value of the concentration parameter as a function  
of the inner slope of the density profile $\beta$ (solid line).  The
value at $\beta=1$ and its $1$-$\sigma$ uncertainty were obtained from
N-body simulations. Concentration for other values of $\beta$ was
obtained by a simple recipe mentioned in the text, and adopting the
same uncertainty in $\log_{10}c_0$.}
\label{fig:c_Nbody}  
\end{figure}  
  
Finally, we use the recipe from Li and Ostriker (2002) to compute
$c_0$ for a GNFW profile given $c_0$ for a pure NFW: we assume that
the ratio $r_{1/2}/r_{200}$ is independent of the density profile
slope, where $r_{1/2}$ is the defined as $M(r<r_{1/2})=1/2
M(r<r_{200})$. We retain the redshift and mass dependence of a GNFW
profile as indicated in Eq.~(\ref{eq:c_Nbody}), as well as the same
uncertainty in $\log_{10}c_0$. Figure~\ref{fig:c_Nbody} shows the mean
value of the parameter $c_0$ and its standard deviation, both as a
function of $\beta$.
  
\subsection{Cross-section for the GNFW profile}  
  
Lensing by GNFW halos has been thoroughly explored by Wyithe et
al. (2001) and Li \& Ostriker (2002), and here we recapitulate the
main results. The lens equation for a spherical symmetric lens is
(Schneider, Ehlers \& Falco 1993)
  
\begin{equation}  
\vec{\beta} = \vec{\theta} - \vec{\alpha}(\vec{\theta})\frac{D_{ls}}{D_{s}}  
\end{equation}  
  
\noindent where $\vec{\beta}$ is the angular location of the source,  
$\vec{\theta}$ the angular location of the lens, and $\vec{\alpha}$ 
the deflection angle\footnote{The $\alpha$ and $\beta$ used in this
subsection are not to be confused with the Schechter function
parameter $\alpha$ and the GNFW profile slope $\beta$ used in the rest
of the paper. Furthermore, note that $\sigma_{\rm GNFW}$ is the
cross-section, while $\sigma$ and $\sigma_*$ refer to the galaxy
velocity dispersion.}.  $D_{ls}$ and $D_{s}$ are the angular
diameter distances between lens and source and observer and source
respectively.  Define $\vec{\xi}$ and $\vec{\eta}$ to be the position
vectors in the lens and source planes respectively, and $x\equiv
\xi/r_s$ and $y \equiv (\eta/r_s)(D_{l}/D_{s})$, where $D_{l} $ is the
angular diameter distance to the lensing object. Then the surface mass
density is given by
  
\begin{equation}  
\Sigma (x) = 2\rho_s r_s \int_0^{\infty}   
(x^2+z^2)^{-\beta/2}\left((x^2+z^2)^{1/2}+1)\right)^{-3+\beta}dz  
\end{equation}  
  
\noindent and the mass by  
  
\begin{equation}  
M(x) = 2\pi r_s^2 \int_0^x x'\Sigma(x')dx'.  
\end{equation}  
  
The deflection angle for a spherically symmetric source is  
  
\begin{equation}  
\alpha(x) = \frac{4GM(x)}{c^2r_s x}.  
\end{equation}  
  
The lens equation then becomes 

\begin{equation}  
y = x - \mu_s\frac{g(x)}{x}  
\end{equation}  
  
\noindent where  
  
\begin{eqnarray}  
g(x)         &\equiv& \frac{M(x)}{4\pi \rho_s r_s^3} \label{eq:g}\\ 
\mu_s        &\equiv& \frac{4 \rho_s r_s}{\Sigma_{crit}} \label{eq:mu_s}\\  
\Sigma_{crit}&\equiv& \frac{c_{\rm light}^2}{4\pi G}\frac{D_{s}}{D_{l}D_{ls}},
\end{eqnarray}  
  
\noindent and $c_{\rm light}$ is the speed of light (to be distinguished 
from the concentration). Multiple images occur for $x$ between $\pm
x_c$, where $x_c$ is the solution of $dy/dx= 0$.  Thus the cross
section for the GNFW lens is
  
\begin{equation}  
\sigma_{\rm GNFW} = \pi\left [y(x_c) r_s \right ]^2.  
\end{equation}

\subsection{GNFW optical depth for lensing}
  
The optical depth for the GNFW lens is completely specified by
properties of the lens, $\beta$ and $c(z)$, the locations of the
lens and source, $z_l$ and $z_s$, and the cosmological abundance of the
lenses. As in the SIS case, we use the Schechter luminosity function
to model the number density of galaxies, together with the
Faber-Jackson relation. The optical depth has the same form as in the
SIS case:
  
\begin{eqnarray}  
\tau(z_s) &=& \int_0^{z_s} dz_l {dD_l\over dz_l}\,(1+z_l)^3\times\nonumber\\  
&&\int_0^\infty dL\,  {d\phi\over dL}(L, z_l)\,  
\sigma_{\rm GNFW}(z_l, L)\,B(z_l, z_s, L).  
\end{eqnarray}    

In order to relate the optical depth to the parameters of a Schechter
luminosity function, it is typical to define a one dimensional
dispersion velocity of a GNFW profile in analogy to that defined for
an SIS galaxy:
  
\begin{equation}  
\sigma^2 = \frac{GM}{2\,r_{200}}  
\end{equation}  

Combined with Eq.~(\ref{eq:M200_1}), this gives the mass as a function
of the dispersion velocity

\begin{equation}  
M = \frac{\sigma^3}{G}\sqrt{\frac{3}{100\pi G\rho_c}}  
\end{equation}  

\noindent This mass then determines $r_s$ 

\begin{equation}  
r_s(z) = \frac{1}{c(z)}\frac{\sqrt{2}}{10}\frac{\sigma}{H(z)}  
\end{equation}  

\noindent which, together with $\rho_s(z)$, Eq.~(\ref{eq:rhos}), 
specifies $\mu_s$, Eq.~(\ref{eq:mu_s}), which is necessary for the
lensing equation.

Finally, we need the magnification bias for the GNFW halos. For the
source objects with the power-law flux distribution, as is the case
with CLASS, this is given by (Li \& Ostriker 2002)

\begin{equation}
B=\frac{2}{3-\beta}A_m^{\beta-1}
\label{eq:magbias}
\end{equation}

\noindent where

\begin{equation}
A_m=\frac{2x_0}{y(x_c)y'(x_0)}.
\label{eq:Am}
\end{equation}

\noindent where prime denotes the derivative  with respect to $x$ and
and $x_0$ is defined by

\begin{equation}
y(x_0)=0. 
\end{equation}

Equation (\ref{eq:Am}) has been adopted from Oguri et al. (2002);
magnification bias defined this way agrees very well with ray-tracing
simulations (C.-P, Ma, private communication). The magnification bias
for GNFW halos is very large, of order a few tens or hundreds.

\section{GNFW profile -- results}
 
\subsection{Dependence on $\beta$}

To compute lensing statistics using the GNFW profile, we need to
supply $c(z)$ and $\beta$. Our main goal here is to {\it determine}
the inner density profile $\beta$, which is a parameter of
considerable interest and to which the lensing statistics are very
sensitive. Therefore, we marginalize over the concentration normalization
$c_0$ and parameters of the Schechter luminosity function.

\begin{figure}  
\includegraphics[height=3.5in, width= 3in, angle=-90]{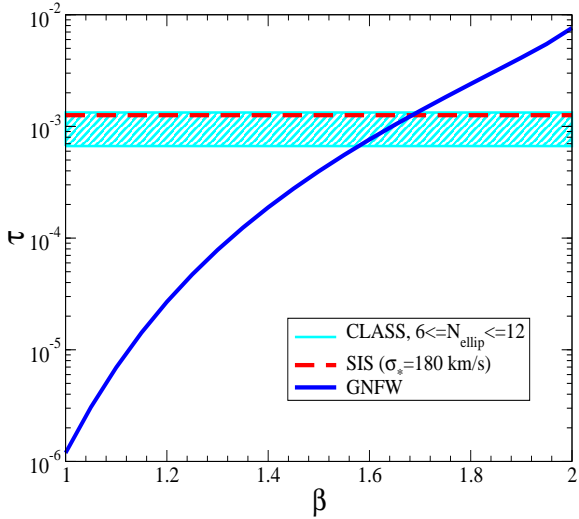}  
\caption{Optical depth vs. the value of the inner density slope, $\beta$  
(rising solid line) for the fiducial values of Schechter function
parameters from Sec.~\ref{sec:SIS_depend}. Also shown is the value
predicted by SIS profile (horizontal dashed line). The optical depth
inferred from CLASS is shown with the shaded region. }
\label{fig:tau_vs_beta}  
\end{figure}

Figure \ref{fig:tau_vs_beta} shows the total optical depth of a GNFW
lens as a function of $\beta$ for $1\leq \beta \leq 2$ and fiducial values of
all other parameters.  Also shown is the value predicted by the SIS
model (horizontal dashed line), also with fiducial values of other parameters, 
as well as the optical depth actually measured by CLASS
(shaded region). (As remarked before, there is no reason that 
the GNFW profile at $\beta=2$ should match the SIS profile case.)
From this figure it is clear that the optical depth is a
strong function of $\beta$. Note too that the lensing cross-section
for $\beta>2$ is formally infinite, although it becomes finite if one
considers configurations in which both images are detectable.  While 
values $\beta>2$ are allowed by our analysis, they are disfavored, 
and for computational reasons we only consider values $\beta\leq 2$.

\begin{figure*}
\includegraphics[height=3.5in, width=2.5in, angle=-90]
	{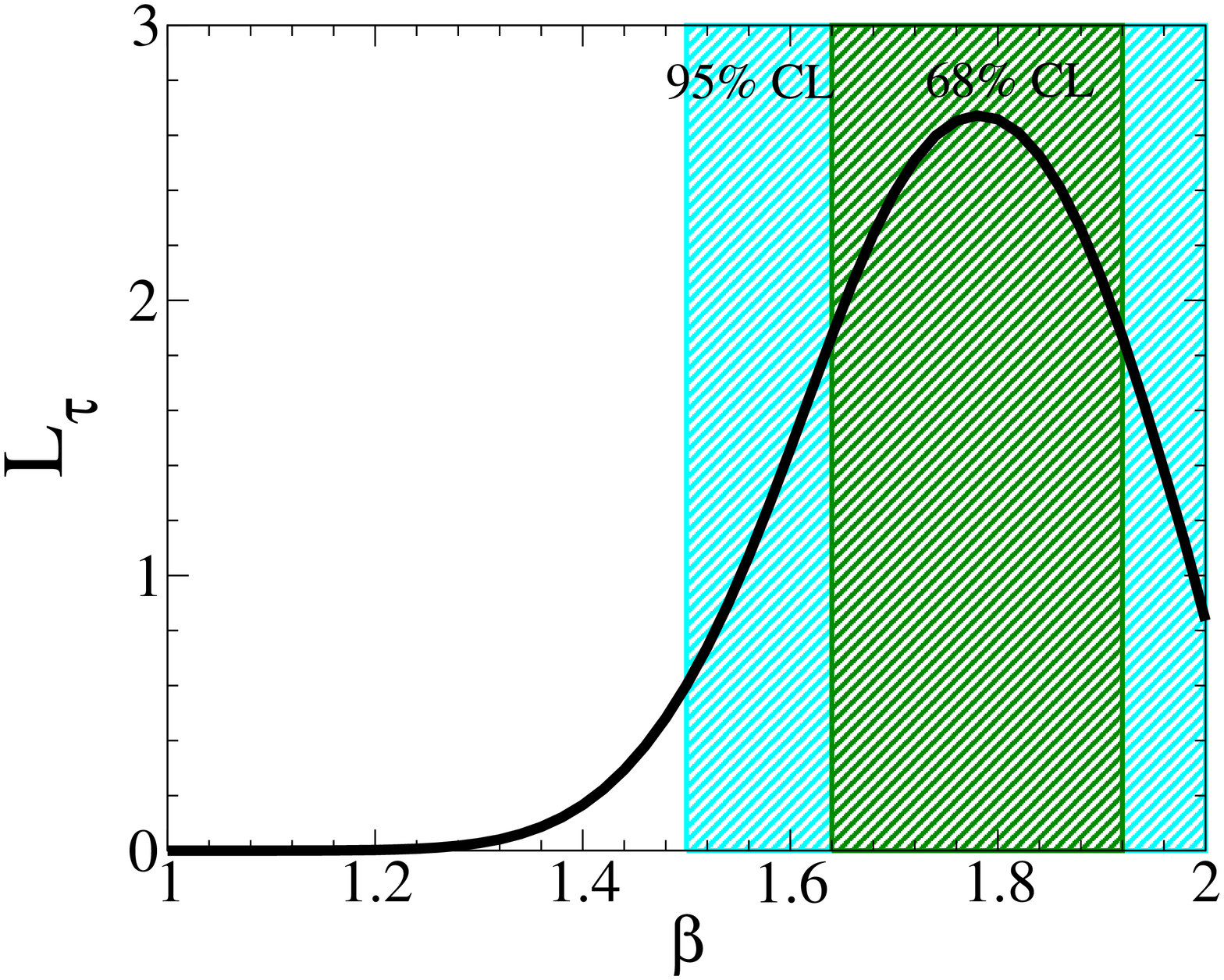}\nobreak\hspace{-0.4cm}
\includegraphics[height=3.5in, width=2.5in, angle=-90]
	{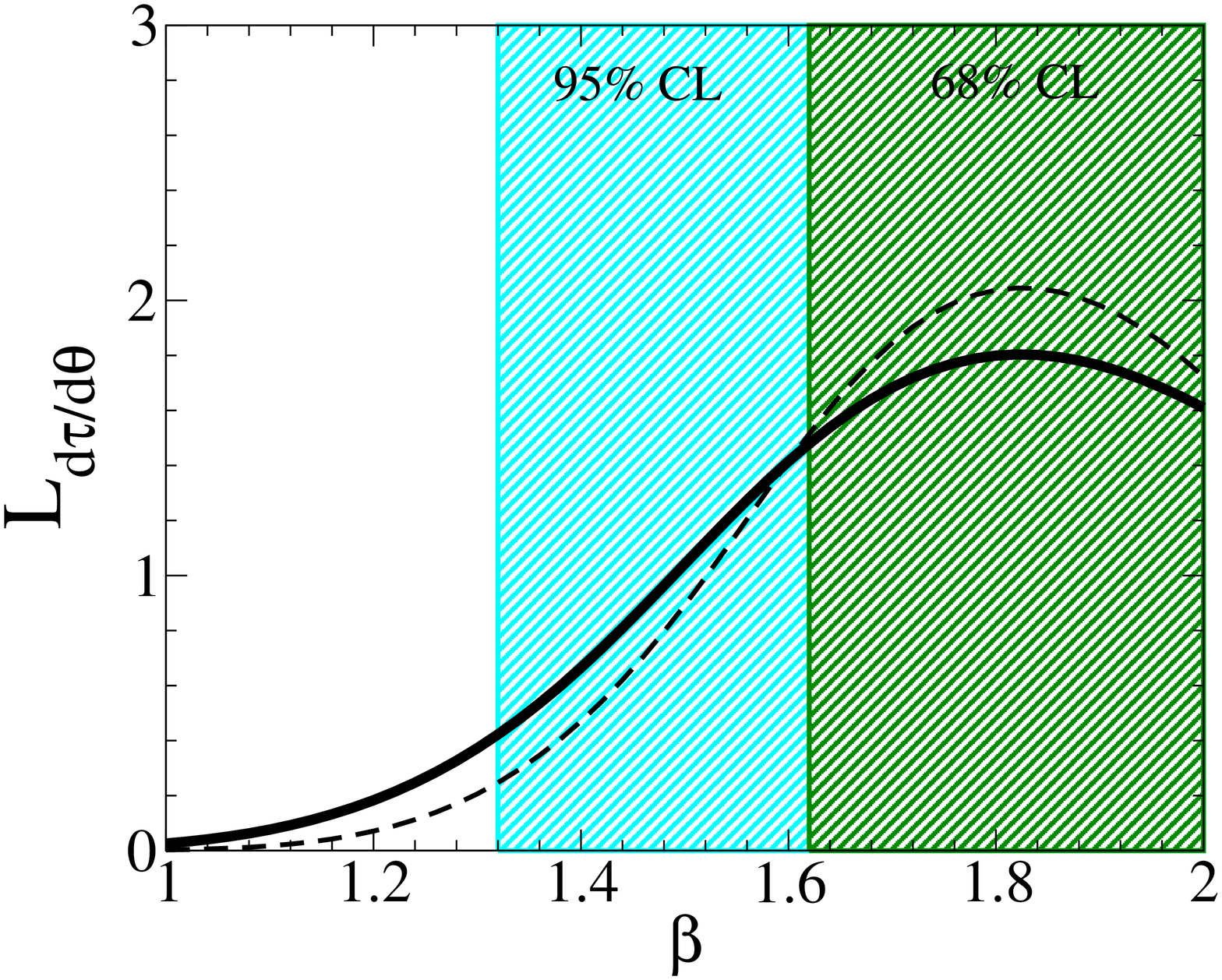}\\[-0.15cm]
\includegraphics[height=3.5in, width=2.5in, angle=-90]
	{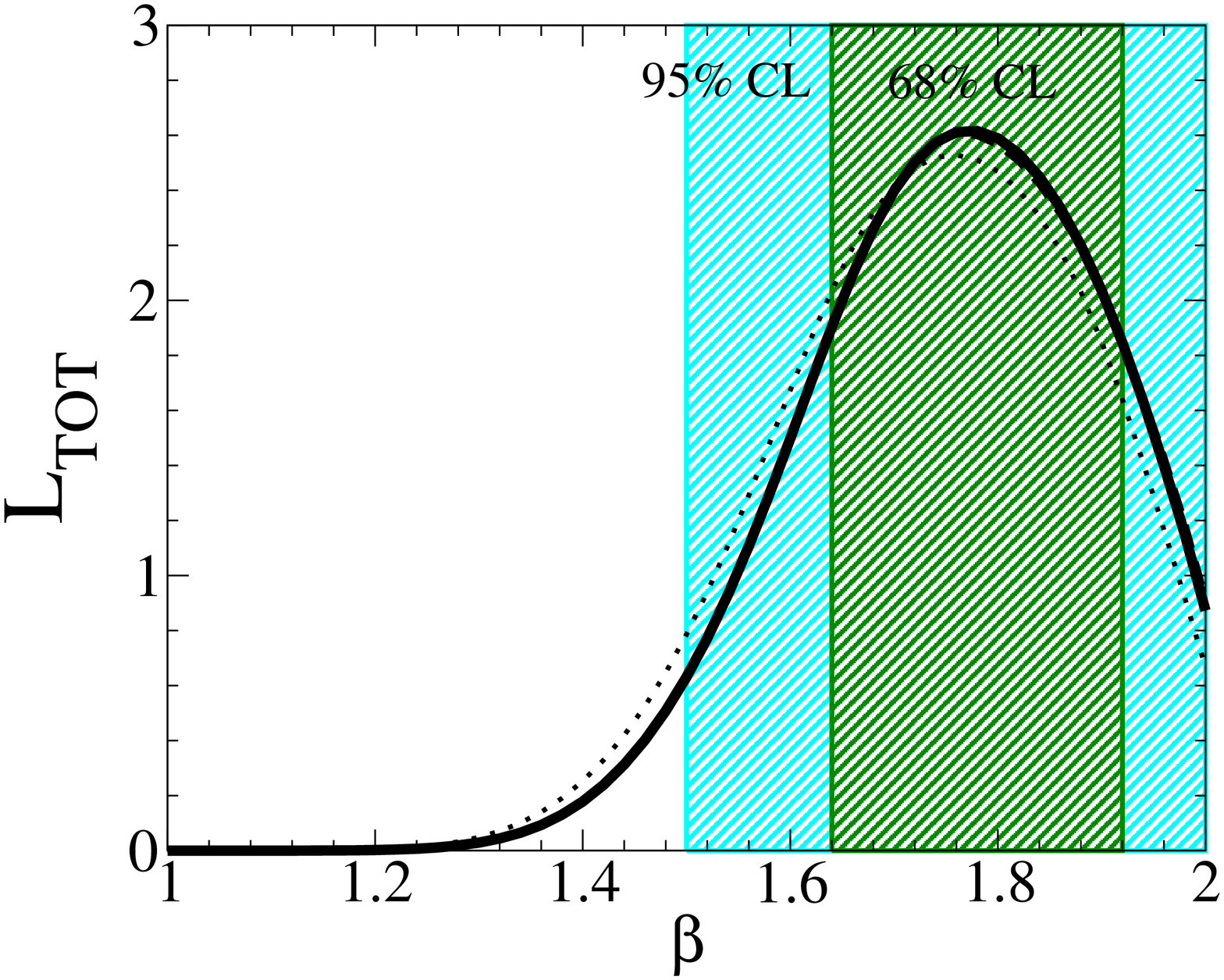}\nobreak\hspace{-0.4cm}
\includegraphics[height=3.5in, width=2.5in, angle=-90]
	{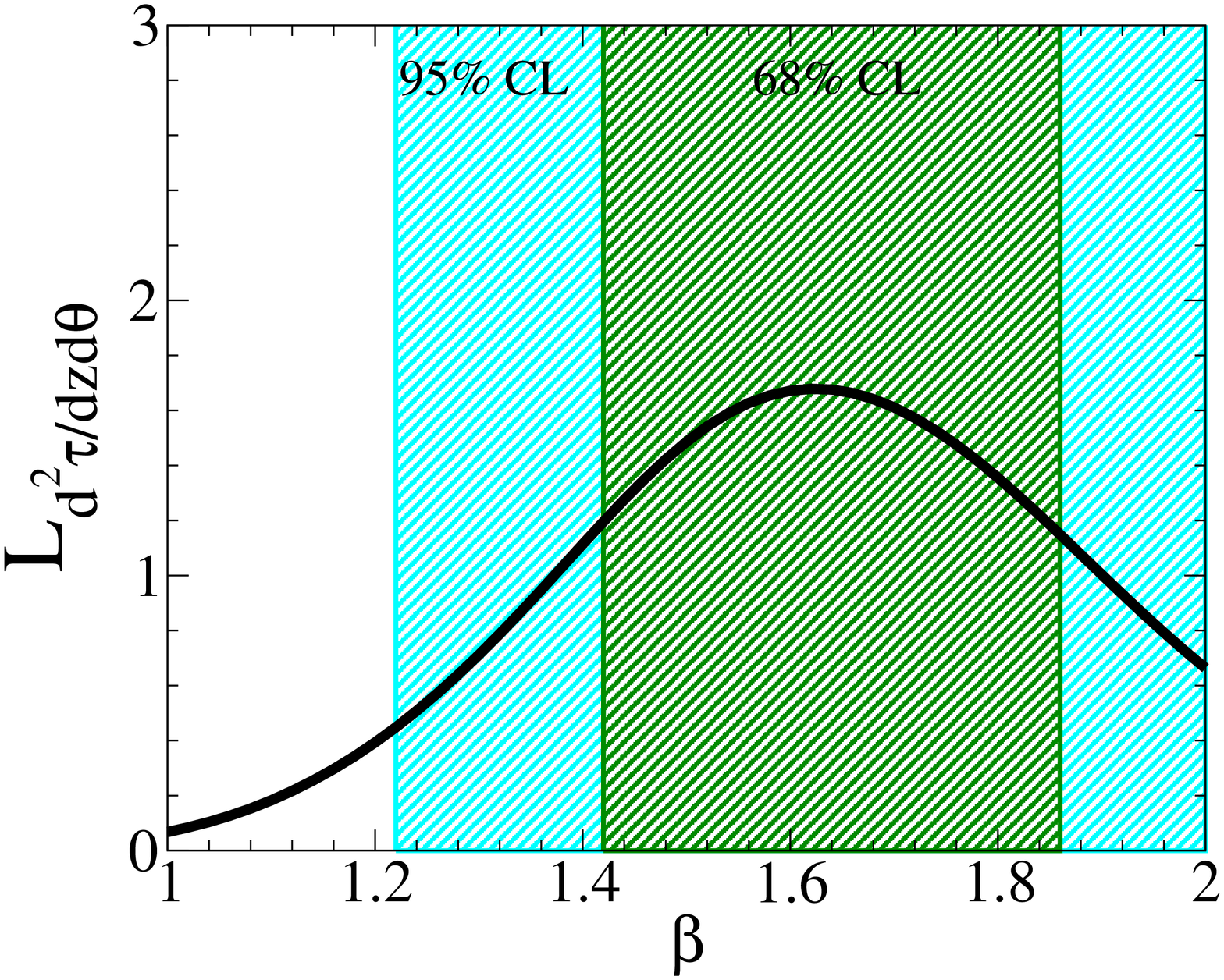}\\
\caption{Constraints on the inner density slope $\beta$, marginalized over
all other relevant parameters. Shown are the likelihood functions for
$\tau$ and $(1/\tau)(d\tau/d\theta)$ (top panels), as well as for the
two combined (lower left panel). The lower right panel shows the
likelihood for the angular and redshift test combined
($(1/\tau)(d^2\tau/dz_l d\theta)$), which we did not use in the
analysis due to the uncertain redshift selection function, but show
for illustration that it is consistent with the other tests. For the
likelihoods using the angular separation test, we show the results
assuming 4 single deflectors confirmed to be ellipticals (solid
lines), and, alternatively, all 9 single deflectors that are not
identified as spirals (dashed lines). Note that the solid and dashed
line in the combined likelihood test esentially overlap. Dotted line
in the combined likelihood test shows the likelihood when $\tau$ and
$(1/\tau)(d^2\tau/dz_l d\theta)$ tests are combined.}
\label{fig:GNFW_constr}
\end{figure*}

\subsection{Parameter choices}  

Our goal is to constrain the density profile $\beta$. We therefore
have to consider how to include a host of other parameters. We adopt
the concentration function $c(z)$ from N-body simulations, using
Eq.~(\ref{eq:c_Nbody}) and choosing $c_0$ with a gaussian prior as
discussed previously. We also need to marginalize over four luminosity
function parameters ($\phi_*$, $\sigma_*$, $\alpha$ and $\gamma$).  We
choose the same ranges for $\alpha$, $\gamma$ and $\phi_*$ as in the
SIS case (see Sec.~\ref{SIS:results}), plus a uniform prior $\sigma_*\in [150,
220]\,{\rm km/s}$, which is indicated by our SIS results.  Remarkably, we
find that interesting constraints on $\beta$ are possible despite
marginalizing over this large parameter space. As before, we assume
the concordance cosmology ($\Omega_M=1-\Omega_{\rm DE}=0.3$; $w=-1$).
  
\subsection{Constraints on $\beta$}  

The resulting constraints on the inner slope of the density profile
are shown in Fig.~\ref{fig:GNFW_constr}.  First, note that the total
optical depth and angular separation tests (top panels) are in good
agreement.  The two tests together, when the likelihood function is
marginalized over other parameters, yield the constraint $1.64\leq
\beta\leq 1.92$ at the 68\% CL and $1.50\leq\beta\leq 2.00$ at the 95\% CL 
(bottom left panel).  As in the SIS case, these results are
insensitive to the exact ranges allowed for the luminosity function
parameters. Moreover, we have checked that the angular separation test
is insensitive to the choice of lens data, i.e. whether we use the 4
single deflectors confirmed to be ellipticals, or all 9 single
deflectors that are not identified as spirals. To be conservative, all
results we quote correspond to the former choice and are represented
by solid lines in Fig.~\ref{fig:GNFW_constr} (for more on this choice,
see Sec.~\ref{sec:data}). We also show the likelihood for the angular
and redshift test combined using the {\it three} elliptical lenses
with complete redshift and angular separation information
($(1/\tau)(d^2\tau/dz_l d\theta)$; bottom right panel), which we did
not use in the analysis due to uncertain systematic effects in the
selection of lens redshifts. It is clear that the combined angular and
redshift test is consistent with the other tests, and combining it
with the $\tau$-test would further strengthen the final constraint on
$\beta$, as shown with the dotted curve in the bottom left panel.

Although the favored slope is significantly steeper that the canonical
NFW $\rho\propto r^{-1}$ profile, it is expected that the shallow NFW
profiles seen in simulations become steeper due to baryonic infall
(e.g. Kochanek \& White 2001). The results of our analysis are in
excellent agreement with such a scenario.  Furthermore, these
constraints are in good agreement with direct modelling of the
observed lenses (Mu\~{n}oz, Kochanek \& Keeton 2001, Cohn et al.\
2001, Treu \& Koopmans 2002, Winn, Rusin \& Kochanek 2002) 
which typically favors a steep, near-isothermal cusp. Finally, the
results are insensitive to the exact values of cosmological
parameters: for example, marginalizing over the plausible values of
the matter density $\Omega_M\in [0.15, 0.40]$ produces negligible
increase of the width of our contours.


\begin{figure}  
\includegraphics[height=3.5in, width= 3in, angle=-90]{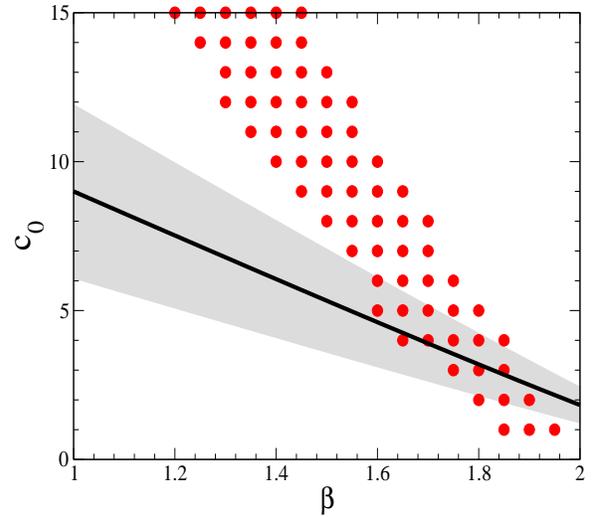}  
\caption{Same as Fig.~\ref{fig:c_Nbody}, but overlapped with the 68\%
CL constraint on $\beta$ and $c_0$ from the $\tau$-test of lensing
statistics (filled circles). We use the total optical depth for the
latter constraint, and retain the redshift and mass dependence of
$c(z)$ as in Eq.~(\ref{eq:c_Nbody}). Note that the region of overlap
coincides roughly with the allowed range of these two parameters based
on the likelihood function. }
\label{fig:beta_c0.joint}  
\end{figure}

We also can constrain the GNFW concentration parameter $c(z)$ and the density
profile slope $\beta$ jointly.  In Fig.~\ref{fig:beta_c0.joint} we
display the N-body determination of the concentration parameter as a
function of $\beta$, and overlay this with our lensing constraint on
$c_0$ vs $\beta$, using the $\tau$-test. For any given $\beta$, we
allow $c_0$ to be a free parameter, and retain the redshift and mass
dependence of $c(z)$ as in Eq.~(\ref{eq:c_Nbody}). Not surprisingly,
the allowed value of $\beta$ reported above coincides with the overlap
region between the N-body result and our lensing constraint. Note,
however, that lensing imposes constraints on the concentration that
are independent of N-body results. In particular, if the galaxies
indeed have  pure NFW ($\beta=1$) profile, lensing statistics
implies that the concentration parameter $c_0$ has to be greater than 15,
which is in conflict with the results of N-body simulations.

\section{Redshift-dependence of the luminosity function?}
\label{sec:evolving} 
  
We mentioned previously that one of the great difficulties with using
gravitational lensing statistics as a probe is that the parameters
that describe the abundance of galaxies can depend on redshift. (In
the GNFW case, the concentration parameter $c(z)$ is allowed to vary
with redshift, as predicted by numerical simulations. ) To make
progress, essentially all authors in the past who wanted to use
lensing statistics assumed that these functions were
redshift-independent.  In particular, one expects that the number
density $\phi_*$ and the characteristic velocity dispersion $\sigma_*$
may be strongly dependent on redshift due to galaxy accretion and
mergers.
\footnote{This situation is reminiscent of that in the
analysis of galaxy surveys, where one needs to know the galaxy-to-mass
bias in order to obtain the distribution of matter from the observed
distribution of galaxies. In the past most authors assumed the bias to
be constant, while it is widely suspected that it depends on scale,
redshift, and galaxy type.}.

Direct constraints on the redshift dependence of the luminosity and
abundance of galaxies are still crude, made difficult by poor
statistics and a variety of systematic effects. Even rough agreement
between various surveys has not been achieved. For example, while the
Canada-France Redshift Survey (CFRS; Lilly et al. 1995), the CNOC2
survey (Lin et al. 1999), and the CADIS survey (Fried et al. 2001) all
observe an increase of $\phi_*$ for early-type galaxies between
redshifts of zero and $z\sim 1$, the Autofib survey (Ellis et
al. 1996) and the numerical simulations by Nagamine et al.(2001)
conclude just the opposite. It is clear that getting the redshift
dependence of number densities and characteristic velocities per
spectral type and their various covariances will take some time.
Keeton (2002) has argued that a variation in $\phi_*$ with $z$ can
cancel out much of the cosmological sensitivity of lensing statistics.
However, we note that this variation alone is probably unrealistic.
At the same time, mergers and accretion will be expected to cause a
variation $\sigma_*$, which will have the opposite effect of a
variation in $\phi_*$ on lensing statistics, and indeed may overwhelm
it.  To accurately account for evolution, it is probably best to match
onto N-body simulations of the galaxy mass function, which in fact
suggest that the number density of galaxies with a specific value of
$\sigma_*$ is relatively constant with $z$ (i.e.\ Bullock et
al. 2001b).

To estimate the maximal possible effect of evolution (assuming an
evolution in the galaxy number density only), we used an SIS profile,
which simplifies calculations. If one then considers a number density
dependence of galaxies as suggested by Lin et al. (1999)

\begin{equation}
\phi_*(z)=\phi_*(0)10^{0.4Pz}
\label{eq:phiz}
\end{equation}

\noindent one can estimate how the results would change for
non-zero values of $P$. For the $\tau$-test, the change is as
expected: for example, for $P=1$ the number density increases by
$\sim$60\% (assuming the average lens redshift is $\sim 0.5$), which
corresponds to the decrease in the favored $\sigma_*$ by $\sim$10\% in
order to preserve agreement with the measured $\tau$,
cf. Eq.~(\ref{eq:depend}). For the $(1/\tau)(d\tau/d\theta)$ test, the
redshift-dependence of $\phi_*$ largely cancels out in the numerator
and denominator of this quantity.  Therefore, as expected, the total
optical depth is more sensitive to the redshift dependence of
$\phi_*$, while the angular distribution of lenses is not.  Again, we
expect that the actual impact of evolution will be much less severe
than that discussed above, because mergers and accretion will tend to
produce a variation in $\sigma_*$ with $z$ that will cancel the effect
of the variation in $\phi_*$.

\section{Conclusions}  

The use of strong gravitational lensing statistics in order to probe
cosmology has a long history.  Nevertheless, the dominant uncertainty
in the predictions of lensing statistics has to do with estimates of
galaxy parameters, not cosmological ones.  Because of the recent
revolutions in observational cosmology that have allowed us to pin
down the basic cosmological parameters with relatively good accuracy,
gravitational lensing statistics now provide us a new opportunity to
probe the structure of galaxies and the trends of galaxy evolution.
Our results represent a first step in this regard.  Nevertheless, it
is quite remarkable that, in spite of the paucity of lensing
statistics at this time, we obtain non-trivial limits on galaxy
properties.  It is also significant that these limits are largely
independent of cosmological uncertainties.  Since we are primarily
interested here in constraining observational galaxy parameters we
used the Schechter luminosity function, which gives the number density
of galaxies in terms of luminosity, rather than the mass function,
which is more relevant for more massive halos associated with clusters
($M\gtrsim 10^{13}M_{\sun}$).

Assuming the SIS density profile, we find that the mean velocity
dispersion for elliptical galaxies is small, with $ 168\, {\rm km/s} <
\sigma_* < 200\, {\rm km/s}$ at 68\% CL, consistent with a number of earlier 
estimates used in lensing analyses (e.g. Chiba \& Yoshii 1999, Cheng
\& Krauss 2001), but significantly smaller than the ``canonical" value
of $225$ km/s often quoted in the literature.  Perhaps more
significantly, assuming the generalized NFW density profile with inner
slope $1\leq\beta\leq 2$, we  constrain $\beta$ to be in the range $ 1.64
\leq \beta\leq 1.92$ at 68\% CL.  This is definitely inconsistent with the
$\beta=1$ slope advocated by N-body simulations for the dark matter
halos profiles; Fig.~\ref{fig:beta_c0.joint} shows that a profile with
$\beta=1$ could produce the observed lensing statistics only with an
unreasonably high concentration ($c_0 > 15$). At the same time, it is
a well-known fact that N-body simulations do not include additional
physics, e.g. the baryonic infall, that makes the inner profiles of
halos and galaxies steeper. Consequently, our result for the density
slope is in good agreement with the expectations, as well as with
similar analyses (e.g. Keeton 2001, Kochanek \& White 2001, Rusin
\& Ma 2001) or direct modelling (Cohn et al. 2001, 
Treu \& Koopmans 2002, Winn, Rusin \& Kochanek 2002). The lack of
high-separation events ($>\!\! 3''$) in JVAS/CLASS has been used to
suggest that there are two populations of halos in the universe
(Keeton 1998, Li \& Ostriker 2002): small-mass galaxy-size halos with
possibly steep density profiles ($\beta\sim 2$), and large-mass halos
with shallow density profiles ($\beta\sim 1$). The former correspond
to the elliptical galaxies we are interested in here, and our results
confirm that a steep slope seems to be required to explain these
events.

There remain some issues that require further exploration.  In
particular, the total optical depth produces a likelihood function
that tends to suggest a slope that is less steep than that favored by
exploring the redshift dependence and angular splitting of lensing
events.  This may be an artifact of our limited statistics, but it
could also signal the need to consider a more complicated, perhaps two
component, galaxy distribution in order to consistently model lensing
events.

Needless to say, the most significant factor not explicitly taken into
account here is a possible redshift evolution of galaxy number density
and velocity dispersion. Very little is known observationally about
the evolution of these quantities beyond $z\sim 0.3$. As we have
discussed in Sec.~\ref{sec:evolving}, there are reasons to believe that
the effects of evolution will not significantly alter the allowed
parameter ranges we have determined here.  

Finally, we note that there is great potential to improve these
constraints as better statistics are obtained using current and future
observational efforts.  In particular, the DEEP2 redshift survey
(Davis et al. 2002) will provide a velocity function for galaxies at
redshift $z\sim 1$, which will allow one to explore the evolution of
galaxy parameters with redshift with a much higher sensitivity than
currently available.

\section*{Acknowledgments}  
  
We would like to thank Joanne Cohn, Marc Davis, Chuck Keeton,
Chung-Pei Ma, Paul Schechter and Risa Wechsler for useful
conversations, and the anonymous referee for numerous useful comments
and suggestions. DH and LMK acknowledge the hospitality of the Kavli
Institute for Theoretical Physics, where this work was completed.  The
work of the CWRU particle astrophysics group is supported by the
Department of Energy.

\end{document}